\documentclass[conference]{IEEEtran}
\IEEEoverridecommandlockouts
\usepackage{cite}
\usepackage{amsmath,amssymb,amsfonts}
\usepackage{algorithmic}
\usepackage{graphicx}
\graphicspath{ {./images/} }
\usepackage{float}

\usepackage{hyperref}
\usepackage{nameref}
\usepackage{cleveref}
\usepackage{todonotes}

\usepackage{textcomp}
\providecommand{\textapprox}{}
\renewcommand{\textapprox}{\raisebox{0.5ex}{\texttildelow}}

\usepackage{xcolor}
\usepackage{hhline}
\def\BibTeX{{\rm B\kern-.05em{\sc i\kern-.025em b}\kern-.08em
    T\kern-.1667em\lower.7ex\hbox{E}\kern-.125emX}}

\usepackage{glossaries}
\newignoredglossary{hidden}
\glsdisablehyper
\newacronym{netsd}{NetSD}{Networked SD card}
\newacronym{cots}{COTS}{Commercial Off-The-Shelf}
\newacronym[longplural={Devices under Test}]{dut}{DuT}{Device under Test}
\newglossaryentry{rag}{
	type=hidden,
	name={remote access gateway}
}
\newacronym{hwifi}{HWIFI}{Hardware-Implemented Fault Injection}
\newacronym{ic}{IC}{Integrated Circuit}
\newacronym{cf}{CF}{CompactFlash}
\newacronym{spi}{SPI}{Serial Peripheral Interface}
\newacronym{iot}{IoT}{Internet of Things}

\begin{document}


\title{NetSD:\\ \huge{Remote Access to Integrated SD Cards of Embedded Devices}}

\author{\IEEEauthorblockN{Valentin Schröter\IEEEauthorrefmark{1}, Arne Boockmeyer\IEEEauthorrefmark{2} and Lukas Pirl\IEEEauthorrefmark{2}}
\IEEEauthorblockA{\textit{Professorship for Operating Systems and Middleware} \\
\textit{Hasso Plattner Institute, University of Potsdam}\\
Potsdam, Germany \\
Email: \IEEEauthorrefmark{1}{valentin.schroeter}@student.hpi.uni-potsdam.de, \IEEEauthorrefmark{2}{\{firstname.lastname\}}@hpi.uni-potsdam.de}
}

\maketitle

\begin{tikzpicture}[remember picture,overlay]
  \node[anchor=south,yshift=10pt] at (current page.south) {\fbox{\parbox{\dimexpr\textwidth-\fboxsep-\fboxrule\relax}{
    \footnotesize \textcopyright 2021 IEEE. Personal use of this
    material is permitted. Permission from IEEE must be obtained for
    all other uses, in any current or future media, including
    reprinting/republishing this material for advertising or
    promotional purposes, creating new collective works, for resale or
    redistribution to servers or lists, or reuse of any copyrighted
    component of this work in other works.
  }}};
\end{tikzpicture}%

\begin{abstract}
Digitalization continuously pervades all areas and the \gls{iot} is still on the rise.
This leads to an increased need for efficiency in the development of embedded devices and systems composed thereof.
Hybrid testbeds are common environments to representatively assess, e.g., hardware-software interaction, interoperability, and scalability.
Although automation is inevitable to achieve efficiency, not all devices offer interfaces to be fully software-controlled.
Most notably, block devices tend to be inaccessible for software outside a \gls{dut}, especially when the latter is in a dysfunctional state.

This paper introduces the \gls{netsd} which enables remote access to removable block devices. 
The proposed system consists of a hardware part, which enables multiplexed access to a block device (e.g., an SD card) and a software part which enables remote access to the block device (e.g., via HTTP or network block device).
\Gls{netsd} thus adds testing and automation possibilities to \glspl{dut} without the need to modify their hard- or software.
During the hardware design, we fund that different SD transfer modes and access profiles (read or write focus) benefit from different pull-up resistor configurations for the data lines.

\end{abstract}

\glsresetall


\section{Introduction}
\label{sec:introduction}

In modern days, all areas of daily life experience shift towards more digitalization.
Systems are increasingly computer-aided and interconnected.
This also affects safety-critical areas, like the railway industry.
Besides the positive effects like increased efficiency and comfort, the aforementioned shift also comes with the challenge of more complexity and thus more potential causes for defects.
This indicates that especially in the area of safety-critical systems, testing the systems under development and their interplay with other systems need to be a first-class concern throughout their whole life cycle.

Especially for railway infrastructure, large-scale testing of the interoperability of digital devices is a relatively novel challenge.
So far, there were only a few players on the infrastructure market selling their products.
For this small amount of devices and interfaces, manual interoperability testing was feasible.
But due to new open standards, like from the EULYNX initiative\footnote{\url{https://www.eulynx.eu/} (accessed 2021-07-07)}, the market opens for new players and their devices.
This circumstance increases the need for automated testing, as the large number of devices, interfaces, inter-dependencies and synergistic effects make manual testing insensible.

A common way to test the interoperability of software together with the underlying hardware is testing in hybrid testbeds.
A testbed like Marvis\cite{beilharz2021towards} provides a representative and scalable execution environment for hybrid tests.
Therefore, techniques like simulations, virtualization, and hardware-in-the-loop (HIL) are used \cite{abbas2020product}.
A device under development becomes the \gls{dut} to interact with a set of other devices, in order to test if the behavior and communication are interoperable.
Together with the traffic simulator SUMO\footnote{\url{https://sumo.dlr.de} (accessed 2021-07-07)}, Marvis can execute railway scenarios, each containing a new configuration and a new set of connected devices to test various workflows for the \gls{dut}.

This requires, that the \gls{dut} is easily reconfigurable (e.g., firmware version, network configuration, software version).
The reality shows, that easy reconfiguration is mostly not supported by embedded systems.
It is rather the contrary, that remote access to the configuration is prevented by the system hardware and software design.
This reduces the risk for attacks but also limits the capabilities for automated testings.

To add the capability of automated configuration for devices, which are by default not able to be reconfigurable in an automated way but contain a block storage device for configuration, we suggest the \gls{netsd}, a remotely accessible block storage device.
The project arose from the use case of improving the automated testing capabilities of a railway axle counter, which stores its configuration data on integrated block storage that is not accessible remotely.
\Gls{netsd} makes the block storage device --- including the possibly contained file systems and files --- remotely accessible without physical access to the device itself.
Besides remote access, \gls{netsd} also enables the injection of faults at hardware level, which opens new possibilities for testing storage hosts.

The following paper is structured as follows: \cref{sec:relatedWork} provides an overview of other solutions of remotely accessible block storage devices.
\Cref{sec:implementation} describes the details about our hardware implementation.
\Cref{sec:hardwareEvaluation} evaluates the performance of the previously described implementation.
\Cref{sec:software} shortly illustrates the software enabling remote access.
\Cref{sec:conclusionFutureWork} summarizes the capabilities of the implementation to fulfill our use case and concludes the paper by giving an overview of future work.

\section{Related Work}  \label{sec:relatedWork} 
Around the year 2014 remote SDs\footnote{ like Toshiba FlashAir, Transcend Wi-Fi, Eye-Fi mobi and ez Share} were popular in the area of photography.
These SDs have WiFi capabilities directly on board to access the camera's file storage.
Thus they would be a valid choice for just transferring data from and to a host device.
However, WiFi SDs have multiple problems:
\begin{itemize}
	\item Remote SDs only work when the host device is powered.
	It would not be possible at all to access the SD when the host device is switched off.
	\item They are running a custom and inaccessible firmware. Modifications would require reverse engineering\footnote{ For example: \\ \url{http://haxit.blogspot.com/2013/08/hacking-transcend-wifi-sd-cards.html} (accessed 2021-07-07)}.
	\item The connection settings like IP addresses, the WiFi SSID, and passwords are stored on the SD itself. For any changes, it is required to remove the SD from the \gls{dut}.
	\item On the other hand, WiFi SDs could not grant exclusive data access nor the guarantee, that there is no data corrupted during the operation of embedded systems.
\end{itemize}

WiFi SDs could be sufficient for specific applications.
However, the option to access data while the device is turned off is already missing for our use case.
Apart from this, we would not have any possibility to expand the functionality regarding advanced testbed features that are presented in \cref{sec:conclusionFutureWork} \nameref{sec:conclusionFutureWork}.

Besides WiFi SDs, to the best of our knowledge, we haven't found any closely related publications concerning direct network-accessible block storage devices.

\section{Implementation}
\label{sec:implementation}

The project goal includes, that \gls{netsd} enables remote access to the block storage without the need for modifying the software of the \gls{dut}.
This aspect of the project is especially important for complex systems like railway infrastructure devices, that are not open source or easily adaptable.
\Gls{netsd} must not interfere with the actual operation in any unintentional way.
Thus, while expanding the SD with remote capabilities, it should appear unchanged and be controllable from the \gls{dut} via the same interface.

This leads to the two relevant aspects of the project:
\begin{itemize}
	\item Implement a device enabling network access to an SD.
	\item Implement a mechanism to enable shared access to the SD from both our \gls{rag} and the \gls{dut} while maintaining the same interface for the \gls{dut}.
\end{itemize}


For simplicity, we limit our system to SDs, although embedded devices might also use other technologies, such as \gls{cf} cards.
However, some embedded devices, like the axle counter from the initial use case of the project, store its configurations on \gls{cf} cards.
Nowadays, SDs appear to be the most common and various adapters to convert between the different interfaces are available, e.g., \gls{cf}-to-SD.
The decision to base \gls{netsd} on SD technology aims to make the project applicable for a wide range of devices --- be it directly or through adapters.
In the following, we use the term \textit{SD} to not only refer to SD cards themselves but also to refer to block storage technologies having adapters to SDs.

\begin{figure}[htbp] 
	\centering
	\includegraphics[width=\columnwidth]{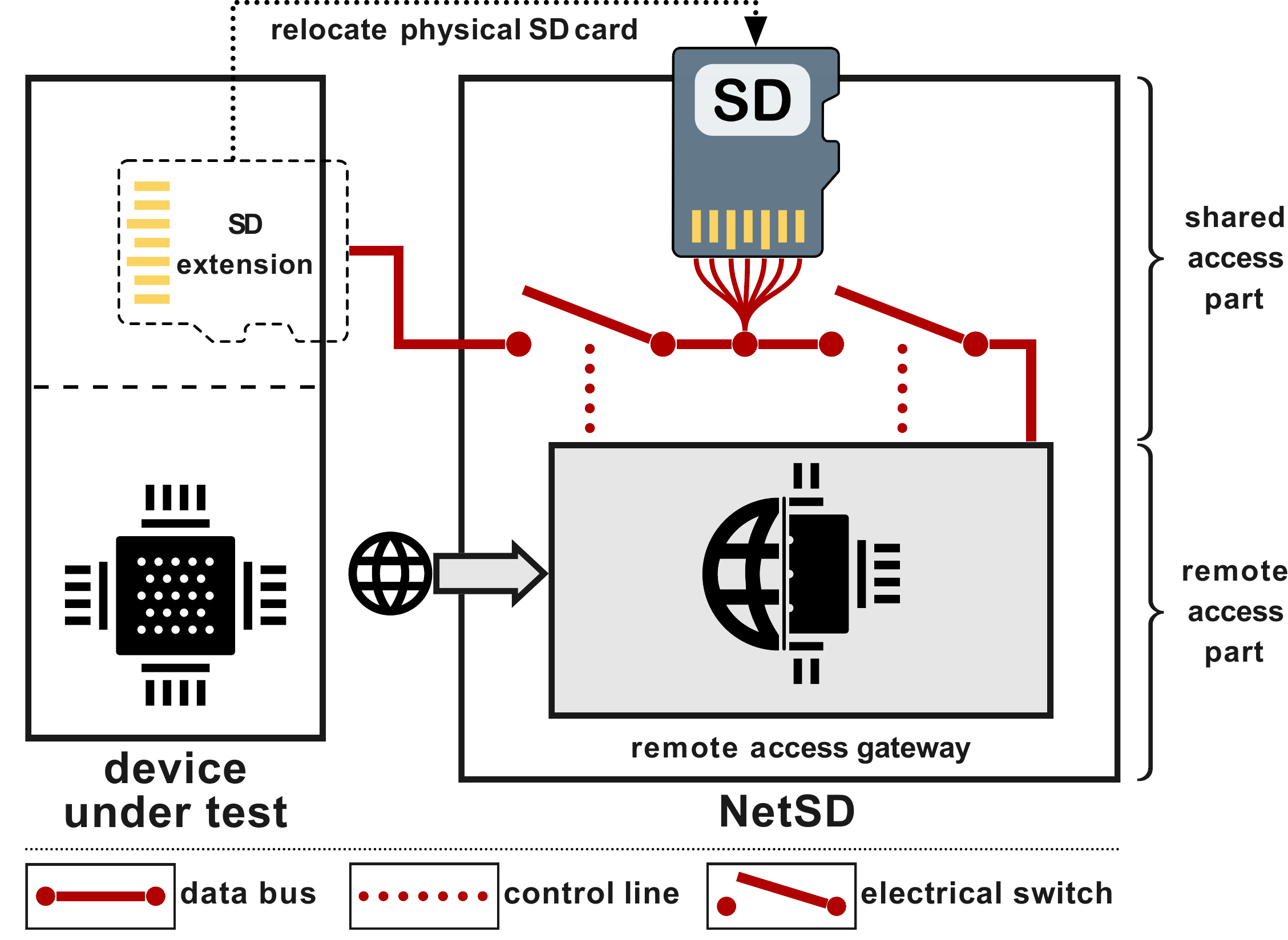}
	\caption{Architecture of a \gls{netsd} setup. The SD of the \gls{dut} is relocated to \gls{netsd} and \gls{netsd} is connected to the \gls{dut} via an extension cable. This enables SD access via the remote access gateway.}
	\label{fig:systemArchitecture}
\end{figure}

\Cref{fig:systemArchitecture} shows the system architecture. Our device, depicted on the right, implements both the remote and the shared access part.

Since \gls{netsd} is intended for development, fault injection, test automation, and alike purposes, neither hardware nor software security are of concern in this project.

\subsection{Remote Access Gateway}
\label{sec:implementation:remoteAccess}

To allow remote access, \gls{netsd} is based on a microcontroller with network capabilities.
Nowadays, there are various microcontrollers and single-board computers offering onboard WiFi, like the Raspberry Pi 4, the ESP32 microcontroller or the wireless STM32 series.
For this project, we decide to use an ESP32 that can connect to existing networks as well as creating an own access point.
As a small-sized, low-level microcontroller with 4~MByte Flash, 320~KByte RAM, and a 160~MHz dual-core processor, it has enough performance to handle simple storage tasks on the SD and also offers fast WiFi performance.
In contrast to a Raspberry Pi, the ESP32 also provides real-time capabilities out of the box which is important for timing aspects of, e.g., storage protocols.

\subsection{Physical Shared Access}
\label{sec:implementation:sharedAccess}

The shared access implementation is the main part of our contribution.
The following problems arise if multiple devices should have simultaneous access to the same SD:
\begin{itemize}
	\item The shared access must support the \gls{dut} and \gls{netsd} having different voltage levels.
	As the SD interface is specified to operate between 1.7~V and 3.6~V\cite[p.~6]{SDSpec} we cannot assume both devices having the same voltage.
	\item Shared access must not lead to short circuit faults.
	While well-implemented SD devices should not endanger system short circuits, custom embedded systems still carry the risk to create short circuits if both devices try to access the SD simultaneously.
	\item Non-exclusive shared access can create data conflicts, as simultaneous operations from both devices can lead to failed transmissions.
\end{itemize}

Due to these problems, a simple electrical connection of all SD data lines between the \gls{dut} and our device would be neither safe nor reliable in operation.
To ensure a functioning shared access, we introduce the \textit{SD switch}.
This part of \gls{netsd} can grant explicit and exclusive access to the SD, either to the \gls{dut} or to our \gls{rag}.
Therefore it is possible to let the \gls{dut} access the SD by default and just revoking the access for a short time if we have to change data by our \gls{rag}.
While the SD switch ensures exclusive access to the SD, it cannot guarantee the \gls{dut} to always access the SD if required.
During access from our device, operations on the SD from the \gls{dut} will fail and have to be repeated.

An SD contains 6 data lines and 2 power lines\cite[p.~12]{SDSpec}.
The SD switch allows to switch on and off bidirectional communication on all data lines as well as switching the power source so that the respective connected device can perform hard resets of the SD.
For switching, multiple electrical components can be considered, while mainly relays, transistors, and MOSFETs are relevant.
\Cref{tab:switchComparison} lists the advantages and disadvantages of each technology.

\begin{table}[htbp]
\caption{\label{tab:switchComparison}Comparison of electrical components for switching signals.}
\begin{tabular}{ |p{.12\columnwidth}||p{.37\columnwidth}|p{.37\columnwidth}|  }
	\hline
	Component & Advantages & Disadvantages\\
	\hhline{|=||=|=|}
	Relays & Easy to design bidirectional switch. Certainty of electrical separation. & Slow control speed. Switching needs a lot of energy. Big component size. \\
	\hline
	Transistors & Fast control speed and high signal transmission frequencies. Small component size.  & Decrease in voltage affects the interface. Complex electrical network to allow bidirectional switch.\\
	\hline
	MOSFETs & Fast control speed switching and high signal transmission frequencies. No changes in signal voltage. Small component size. & Complex electrical network to allow bidirectional switch.\\
	\hline
\end{tabular}
\end{table}

The simplest way to implement an electrical switch would be to use relays.
However, relays as mechanical components are comparatively large, which would accumulate for six data lines per connected device.
A MOSFET on the other side is a non-mechanical, cheap and small-sized component.
An important requirement for switching bidirectional interface lines is the consistency of the voltage levels.
In contrast to a normal transistor, a MOSFET exactly offers this voltage consistency.
That leads to using a MOSFET as the preferred choice for the SD switch basis.
Nevertheless, to ensure a safe electrical separation if the SD switch should disconnect a single device, multiple MOSFETs and other components must be combined in a complex electrical circuit.
But, as analog switching is a default task in electronics, there exist \glspl{ic} offering the desired bidirectional switching functionality with safe electrical separation.
For this project, we chose the \textit{74LVC1G66-Q100} circuit from Nexperia\footnote{ See \url{https://www.nexperia.com/products/automotive-qualified-products-aec-q100-q101/automotive-logic/switches-multiplexers-de-multiplexers/series/74LVC1G66-Q100.html} (accessed 2021-07-07) for Product specifications. Most other analog switches also from other companies would work too.}.

\begin{figure}[htbp]
	\includegraphics[width=\columnwidth]{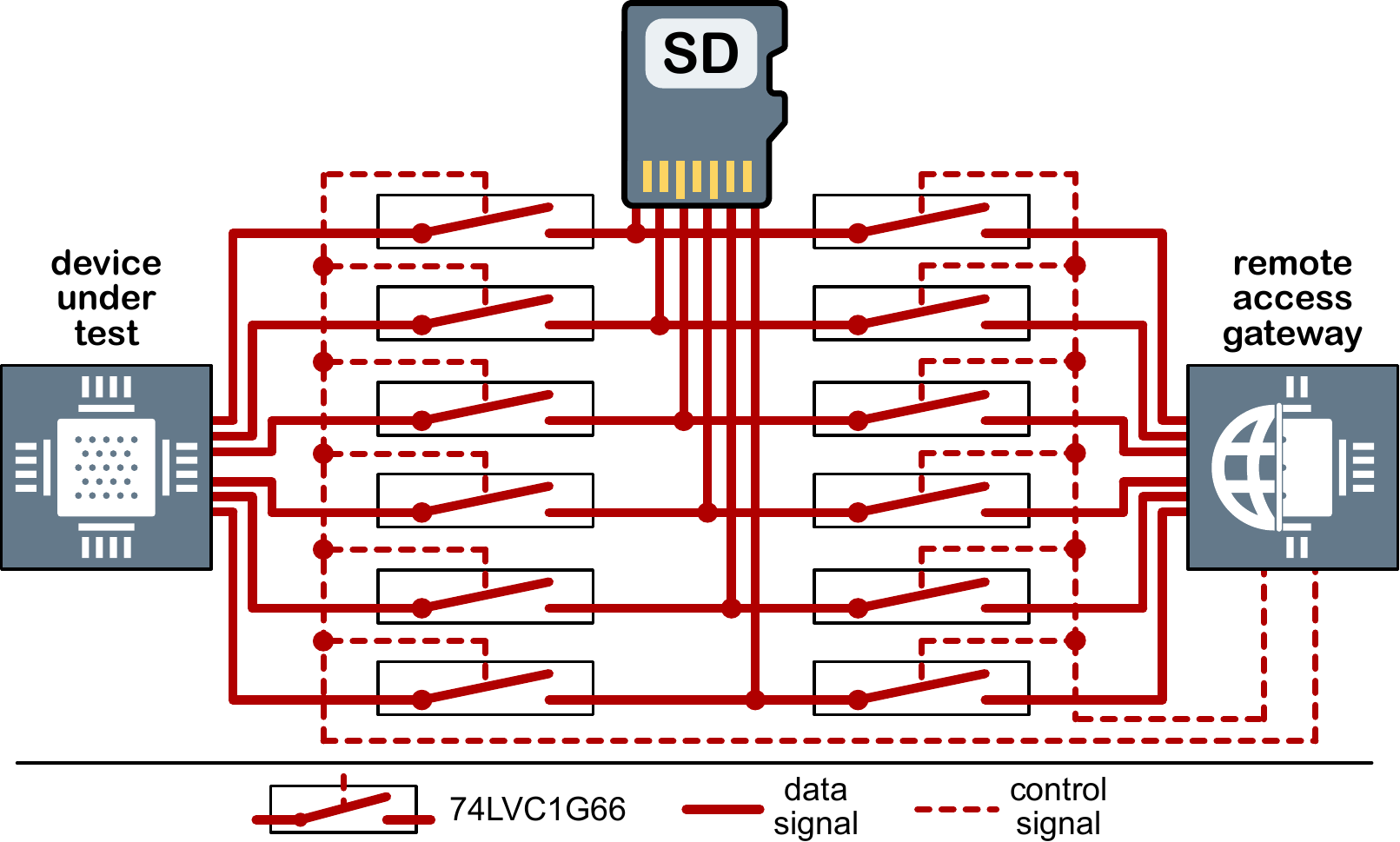}
	\caption{Simplified circuit diagram. Each SD data line is connected to both devices via analog switching \glspl{ic}. It can thus be controlled to which device the SD is connected.}
	\label{fig:switchSchematic}
\end{figure}

\Cref{fig:switchSchematic} depicts how analog switches are used to ensure electrical separation for both devices.
All signals of the SD are connected to two sets of switches, one set for the \gls{dut} on the right and one set for our \gls{rag} on the left.
Both switch sets contain a single switch for each of the six data lines\footnote{see \cref{fig:SignalOrder} for an overview of the SD data lines} from the SD.
All switches are controllable by our \gls{rag}, which takes care of which side currently has granted access to the SD.

\subsection{SD extender cable}  \label{sec:implementation:extenderCable}

\begin{figure}[htbp]
	\begin{center}
		\includegraphics[width=0.7\columnwidth]{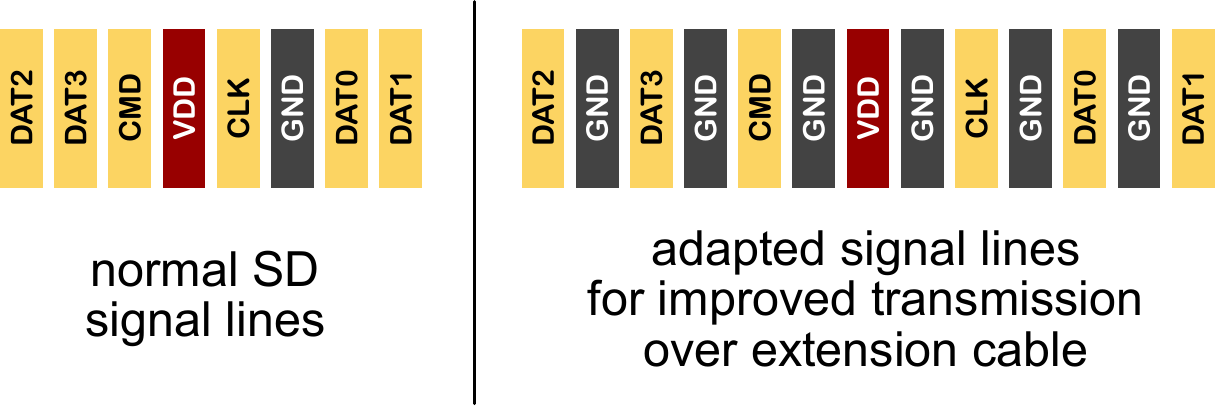}
		\caption{\textbf{Left:} Connection lines of an SD. \textbf{Right:} Connection lines of the extension cable, where a GND line is placed between each data line of the extension cable to avoid crosstalk.}
		\label{fig:SignalOrder}
	\end{center}
\end{figure}

To relocate the physical SD to \gls{netsd}, an extension cable is used, that simulates the SD in the actual slot of the \gls{dut}.
The extension cable uses crosstalk avoiding signal arrangement, which can be seen on the right side of \cref{fig:SignalOrder}.
This arrangement inserts a GND line between each data line, minimizing the mutual influence of signals.

For low-frequency signals (around 10~MHz) it is possible to use the exact same arrangement as the contacts from the SD, which can be seen on the left side of \cref{fig:SignalOrder}.
However, since the native SD interface is specified to transfer data with up to 208~MHz \cite[p.~1]{SDSpec} these direct adjacency of signal lines leads to interference in the transmission.
Thus the crosstalk avoiding signal is important to meet tighter timings and allow reliable transmissions from host devices to the SD.

\subsection{Software}
\label{sec:software}

To enable accessing data on SD via our \gls{rag} with software, \gls{netsd} needs to offer software interfaces to the network.
We implemented two different software interfaces for remote access which are described below.
The ESP32 itself can communicate with the SD through an \gls{spi} and through the native SD interface\cite[p.~12 ff.]{SDSpec}.
Both interfaces allow unrestricted access to the SD.

\subsubsection{REST API}:
The first method is a simple REST API.
The API offers request endpoints the cover all basic file level operations.
This enables simple and platform-independent communication with the connected SD.


\subsubsection{Network block device}:
The second method for remote access is the Network Block Device (NBD) protocol.
This network protocol allows Linux systems to mount remote storage devices as virtual block devices.
Once mounted, an SD can be accessed like normal other mounted devices at a block level.
Therefore using NBD we can work with the remote SD like a normal directory in our local file system, but also flash full images to the SD.
One drawback of NBD is the dependence on using Linux as the operating system.
However, NBD is also available on Windows Subsystem for Linux (WSL).
Furthermore, the data can also be made available to other platforms via protocols such as WebDAV.

\subsubsection{System flow chart}
\begin{figure}[htbp]
	\begin{center}
		\includegraphics[width=0.95\columnwidth]{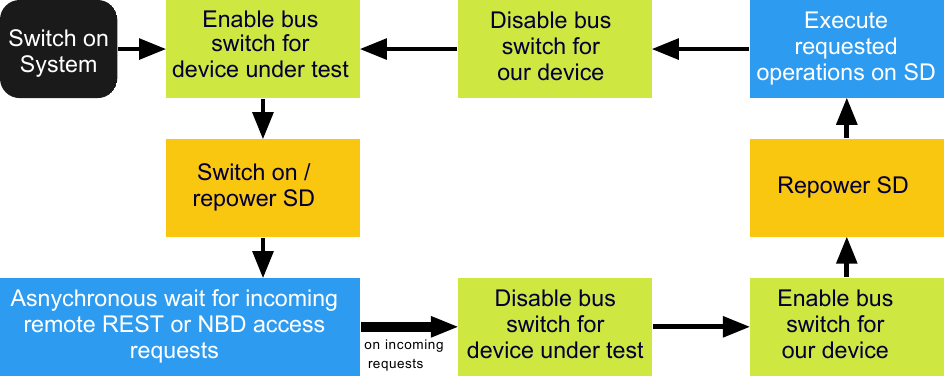}
		\caption{Flowchart of normal operation of \gls{netsd}. In normal operation, only one device is given explicit access to the SD at a time.}
		\label{fig:flowchart}
	\end{center}
\end{figure}

:
\Cref{fig:flowchart} illustrates the normal operation process that allows exclusive access to the block storage.
By default, the \gls{dut} has access to the SD.
Our system is waiting continuously for incoming requests.
On a request, the access to the SD is switched to our \gls{rag}, enabling the execution of requested operations.
After these operations are finished, the access is switched back to the \gls{dut}.
Between each change in access to the SD, it is repowered to allow reinitialization to the communication interface.
This hard reset would not be necessary if concurrent access is implemented like described in \cref{sec:conclusionFutureWork} \nameref{sec:conclusionFutureWork}.
For the project's use case with the axle counter, the device itself will not be turned on until corresponding operations are executed by our device, since the axle counter only read configuration files when starting it.
This procedure however is implemented by the main testbed application that embeds \gls{netsd}.

\section{Evaluation of the hardware prototype}
\label{sec:hardwareEvaluation}

A switch circuit between the physical SD and an accessing device inevitably influences the electrical properties of the data lines.
Especially for \glspl{dut} with fast data transfer requirements, this influence must be as low as possible.
This section presents the hardware quality for the following test conditions:
\begin{itemize}
	\item \textbf{SD}: Intenso microSDHC, 8~GByte, Class 10 (min. 20~MByte/s read throughput, min. 12~MByte/s write throughput).
	\item \textbf{Test tool}: MiniTool Partition Wizard\footnote{ \url{https://www.minitool.com/partition-manager/partition-wizard-home.html} (accessed 2021-07-07)} on Windows 10 operating system testing with different data block sizes.
	Since the evaluation does not test the remote data throughput, but the signal quality for a DuT connected to an SD via the switching circuit, no NBD is required.
	\item \textbf{Baseline test}: Windows 10 computer as \gls{dut} with the SD directly plugged in, also without an extension cable.
	\item \textbf{Test of our SD switch}: Windows 10 computer as \gls{dut} with access to the SD through our SD switch.
	\item \textbf{SD switch specification}: 74LVC1G66 as switch ICs, 48cm long extension cable with crosstalk avoiding signal arrangement, and two different pull-up resistor configurations for the hardware (see below).
\end{itemize}

\begin{figure}[htbp]
	\begin{center}
		\includegraphics[width=0.8\columnwidth]{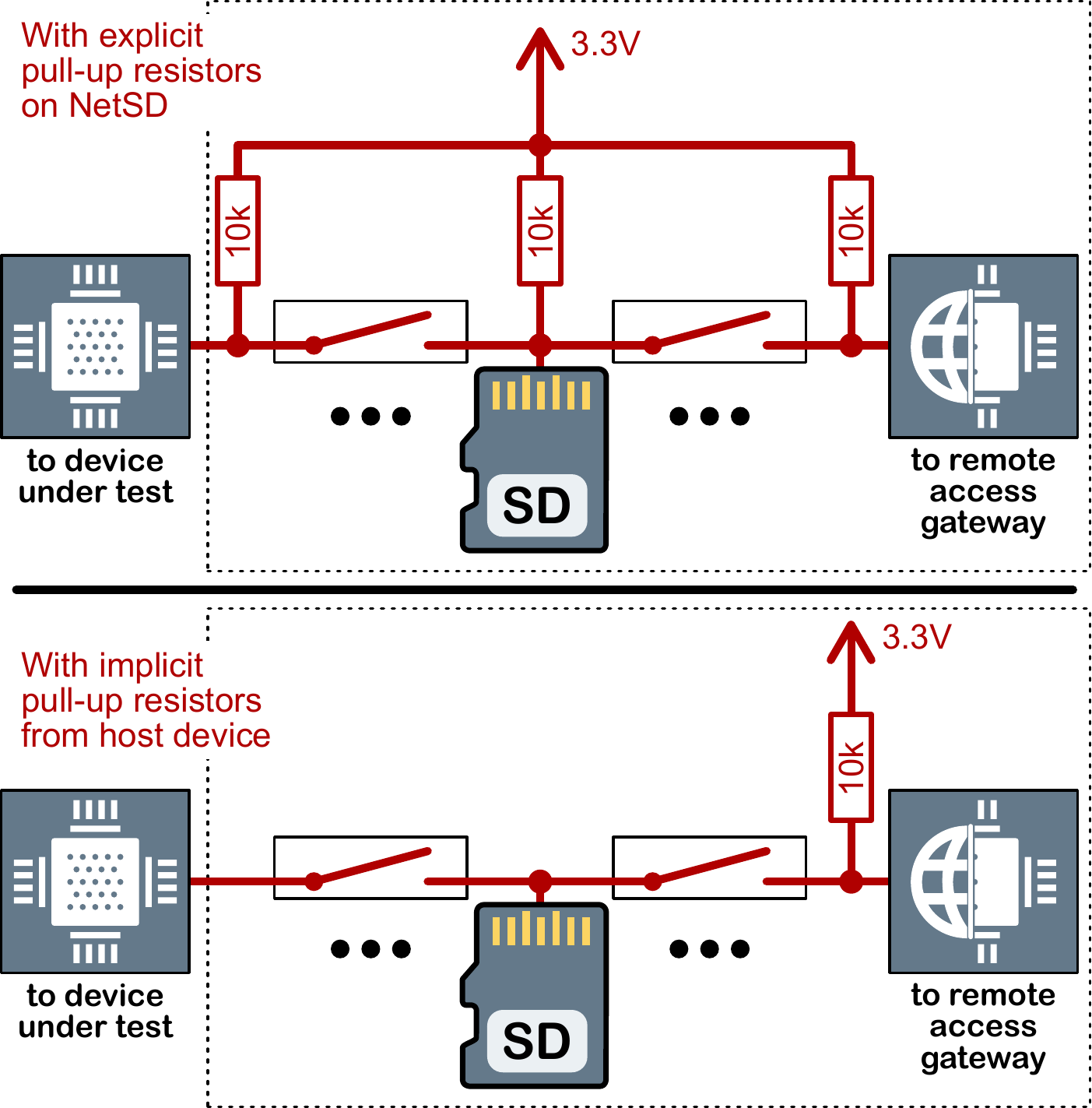}
		\caption{Different hardware setups for data lines on \gls{netsd}. \textbf{Top:} Explicit pull-ups are added on all sides of the analog switches for steeper signal edges. \textbf{Bottom:} No explicit pull-ups are added on side of the \gls{dut}, which provides itself integrated pull-ups.}
		\label{fig:pullupTest}
	\end{center}
\end{figure}

\begin{figure}[htbp]
	\begin{center}
		\includegraphics[width=0.9\columnwidth]{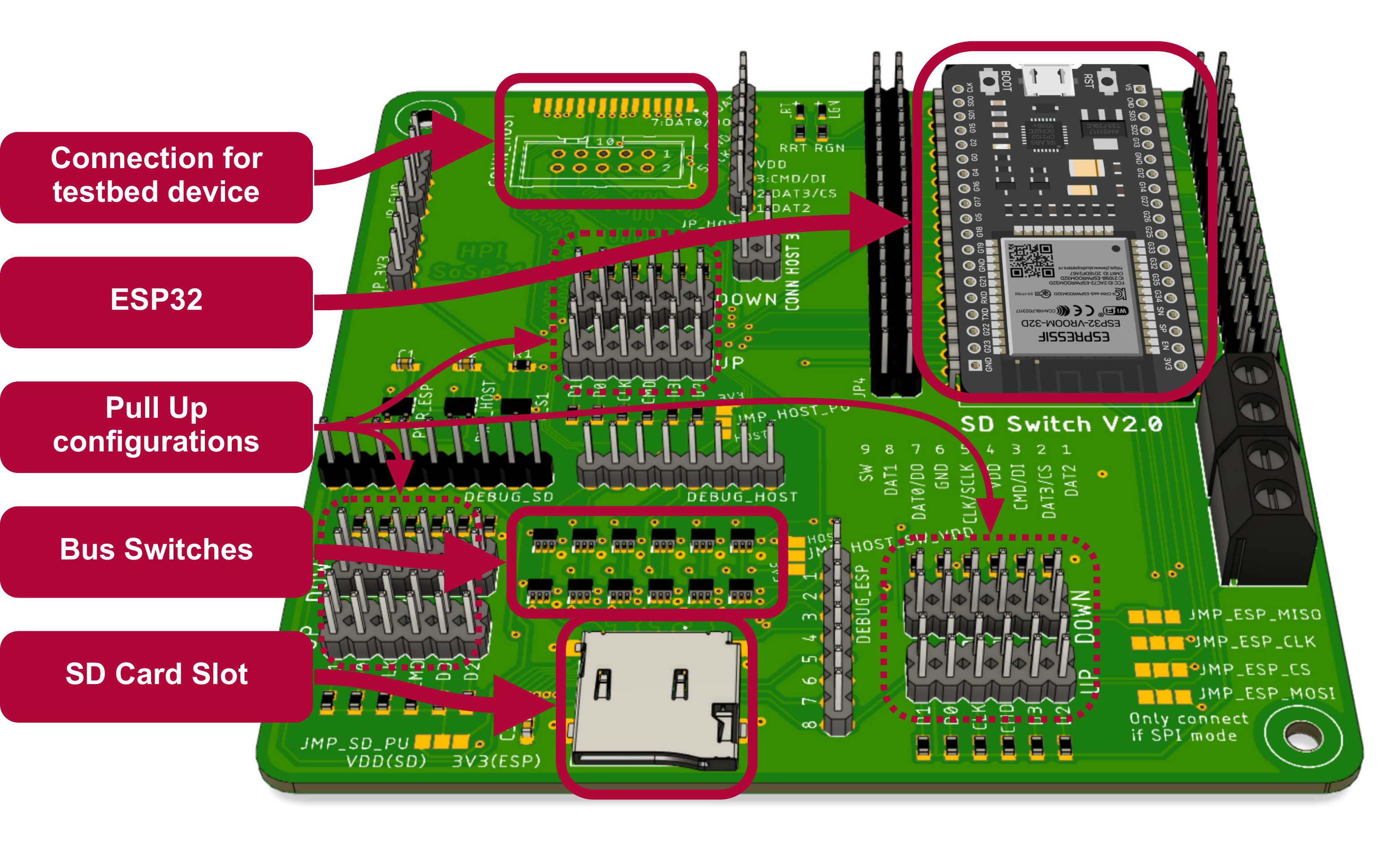}
		\caption{The designed evaluation board for the project. The board focused on offering various debug and configuration possibilities.}
		\label{fig:PCBwith_74LVC1G66}
	\end{center}
\end{figure}

SD interfaces operate with open-drain I/Os\footnote{
	Open-drain I/Os do not actively set both signal states (low and high voltage levels for digital 0 and 1).
	In the case of SDs, signal lines are only actively drawn to GND for digital 0.
	For digital 1, the signal line voltage is not actively set to a high level, but just released to a default state.
}.
These open-drain I/Os require a default voltage level on the signal line induced by the usage of pull-up resistors.
Normally, pull-up resistors are present on the side of the host device.
However, since the physical SD is relocated by an extension cable and the switch \glspl{ic} into \gls{netsd}, these host pull-up resistors are not sufficient to guarantee steep signal edges.
This consideration leads to two different hardware configurations, either by just using the already existent pull-up resistors of the host device (visualized on the bottom of \cref{fig:pullupTest}) or by explicitly adding pull-up resistors on \gls{netsd} to improve signal quality (visualized on the top of \cref{fig:pullupTest}).
To test and debug these different hardware configurations we created an evaluation board that is visible in \cref{fig:PCBwith_74LVC1G66}.

\begin{figure}[htbp]
	\begin{center}
		\includegraphics[width=1.0\columnwidth]{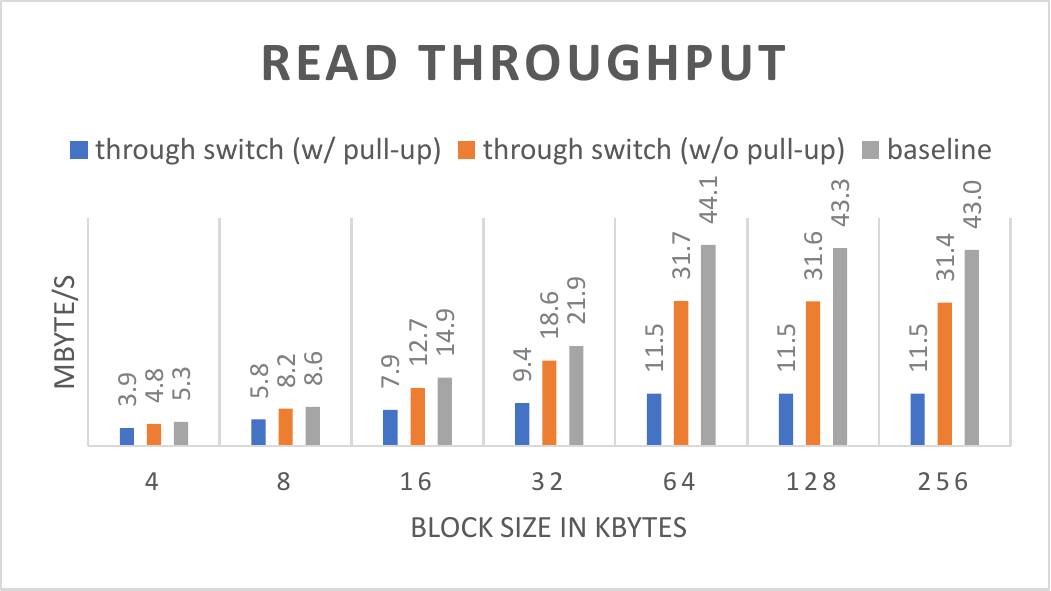}
		\caption{Read throughput with different block sizes and hardware configurations. Throughput increases up until a block size of 64~KByte, where it is \textapprox 3 times higher without explicit pull-up resistors than with and \textapprox 72\% of the baseline throughput.}
		\label{fig:readSpeedTest}
	\end{center}
\end{figure}

\begin{figure}[htbp]
	\begin{center}
		\includegraphics[width=1.0\columnwidth]{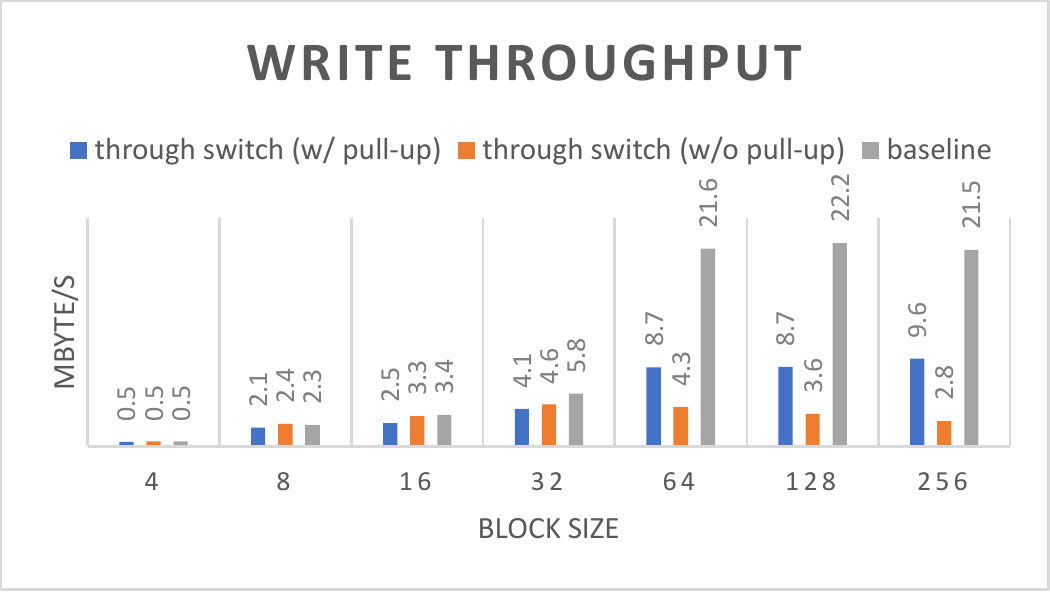}
		\caption{Write throughput with different block sizes and hardware configurations. Throughput increases up until a block size of 64~KByte, where it is \textapprox 2 times higher with explicit pull-up resistors than without and \textapprox 40\% of the baseline throughput.}
		\label{fig:writeSpeedTest}
	\end{center}
\end{figure}

The first test evaluates the read throughput of the SD.
Therefore we compare the read throughput through our switch \textit{with} and \textit{without} an explicitly added pull-up resistor with the baseline read throughput, where the SD is plugged directly into the host device. \Cref{fig:readSpeedTest} visualizes the following read throughput results:

\begin{itemize}
	\item The read throughput generally increases with larger block sizes (with a cap at \textgreater{}64~KByte block size).
	\item The read throughput without pull-up resistors is proportional to the baseline throughput.
	\item The read throughput with pull-up resistors is 30\% to 65\% worse than the read throughput without pull-up resistors.
\end{itemize}

The read throughput test especially shows better performance without explicitly added pull-up resistors.
This can be explained with different bus speed modes of the SD.
The normal speed mode operates at 3.3~V, the same as the supply voltage, which is used for the pull-up resistors too.
However, if both the host device and the SD support the UHS (ultra high speed) mode, the bus operates at 1.8~V.
Without explicit pull-up resistors, this UHS bus mode is automatically selected, resulting in much higher read throughput which is decreased only by the actual hardware delay (extension cable and switch propagation delay) compared to the baseline.
On the other side, if pull-up resistors at 3.3~V are added, the UHS mode cannot be selected, resulting in lower read throughput.

The second test evaluates the write throughput of the SD.
Again, we compare the write throughput through our switch \textit{with} and \textit{without} an explicitly added pull-up resistor with the baseline write throughput.
\Cref{fig:writeSpeedTest} visualizes the following write throughput results:
\begin{itemize}
	\item The write throughput generally increases with larger block sizes (with a cap at \textgreater{}64~KByte block size).
	\item The write throughput without pull-up resistors caps at 32~KByte block size and decreases afterward.
	\item The write throughput with pull-up resistors is better than without pull-ups at block sizes greater than 32~KByte.
\end{itemize}

The write throughput test shows an opposite behavior.
Up to a block size of 32~KByte, the throughput without pull-up resistors is slightly better, but afterward, the write throughput with pull-up resistors becomes much better.
This can be explained by the initial thought of signal degradation if no additional pull-up resistors are installed.
Without pull-up resistors, the UHS mode at 1.8~V is used.
At this low voltage, signal edges become unclean through the long extension cable.
Especially at higher block sizes, this degraded signal quality leads to a higher rate of transmission errors.
These failures result in the data rate cap at a block size of 32~KByte.
On the other side, even though the host does not use the UHS mode if explicit pull-up resistors are added, the signal quality becomes much better, allowing higher write throughput.

In summary, different hardware configurations can take into consideration for different settings.
If the \gls{dut} does not support the UHS mode at 1.8~V, you should add 3.3~V pull-up resistors to improve overall signal quality.
However, if the \gls{dut} supports the UHS mode, the main operation mode is decisive for the decision, whether to use or not to use pull-up resistors.
If primarily read operations are executed, omit the pull-up resistors to achieve higher read throughput.
If primarily write operations are executed, use the pull-up resistors to enforce the 3.3~V mode to achieve higher write throughput.

\section{Conclusion and Future Work}  \label{sec:conclusionFutureWork}

\gls{netsd} has been developed to fulfill our mentioned use case with the axle counter.
The concept proved to be useful and further, \gls{netsd} can aid development practices and test automation of embedded systems in general.

\subsection{Remote access for general purposes}
Any embedded device using an SD (or block storage device with a corresponding adapter available) but not providing remote access to that storage device could be extended by \gls{netsd}.
As explained above, \gls{netsd} is not intended to replace existing remote technologies that are integrated into systems.
\Gls{netsd} rather aims to expand existing systems that are not capable of remote storage access, without changing hardware or software of such systems.

\subsection{Storage multiplexing for embedded systems}
The introduced switch part of \gls{netsd} allowing shared access to the SD could serve as a basis to multiplex a single block storage device between two or more hosts.
Multiple bus switches (one set of switches for each device) would allow safe access to the same physical SD for all connected systems.
This could be used to accumulate data from multiple hosts or to load the same data (e.g., configurations) to multiple hosts.

\subsection{Fault Injection}

The capabilities of the \gls{netsd} hardware prototype offer unprecedented prospects for software-controllable \gls{hwifi}\cite{yuste2003non}.
The forceful insertion (\textit{injection}) of suspected error causes (\textit{faults}) is a well-established approach in testing hardware and --- increasingly ---  software.
Having control over all data lines independently can be leveraged to test hardware (e.g., SDs, host controllers) and software (e.g., device and file system drivers).
Single data lines can be disconnected for short amounts of time to test hardware and software fault tolerance (e.g., error-correcting codes).
Our hardware implementation also allows for modification of the signals on the data lines, such as replaying or corrupting data.
Experiments can thus not only assess crash fault models, but more complex fault models, including computation, omission and timing faults \cite{cristian95atomic}.
It can be suspected, that with more general error classes, more hardware and software misbehavior can be identified.

\subsection{Future work}  \label{sec:futureWork}

The original concept of shared access in \gls{netsd} implied mutually exclusive access: if one device has access to the SD, the other does not.
Originally, this also implied a hard reset to reset the interface settings of the SD after each access switch.


%

With knowledge about the host device and the use of the same interface settings, this principle could be expanded for enabling parallel access.
Instead of switching all lines (incl. the power lines) after switching to the other host device, the system could just switch the data lines without resetting the SD.
However, besides the need to have the same interface settings for both host devices, this method can be expected to only work in highly controllable environments.
E.g., since commonly used file systems do not support parallel access at block level, this approach carries the risk of data corruption.

Instead of implementing parallel access to the SD on a hardware level, it could be realized in software to circumvent the aforementioned shortcomings.
For such a \textit{software SD} to work, a microcontroller could emulate an SD interface via its GPIO lines.
The microcontroller would have exclusive access to the physical SD, so it can safely handle parallel requests from the \gls{dut} and the \gls{rag}.
The challenges with this approach are the implementation of the SD protocol stack and a microcontroller offering IO capabilities fast enough to create the illusion for the hosts to communicate with a physical SD.
Since the SD interface is specified to transfer data at up to 208~MHz \cite[p.~1]{SDSpec} the currently selected ESP32 microcontroller would be too slow for higher frequency modes and therefore not suitable for this approach.
As an extension, the physical SD could be replaced with a network share.

\section*{Acknowledgment}
The authors would like to thank Tobias Zagorni for his active development and testing of \gls{netsd}.


\bibliographystyle{unsrt}
\bibliography{refs}

\end{document}